\documentclass[aps,prl,a4paper
               ,amsmath,amssymb
               ,preprint,showpacs
               ]{revtex4-1}

\usepackage{dcolumn}

\newcommand{\half}{\frac{1}{2}}

\begin{document}
\title{Collective states of interacting $D(D_3)$ non-Abelian anyons}  

\author{Peter E. Finch}
\author{Holger Frahm}

\affiliation{%
Institut f\"ur Theoretische Physik, Leibniz Universit\"at Hannover,
Appelstra\ss{}e 2, 30167 Hannover, Germany}


\begin{abstract}
  We study the finite size spectrum of integrable quantum chains of
  interacting non-Abelian anyons constructed using the Drinfeld double of the
  dihedral group $D_3$.  The gapless low energy modes are identified as the
  direct product of two conformal field theories which can be decomposed
  according to the residual symmetries of the chains subject to periodic
  boundary conditions.
\end{abstract}

\maketitle

%
%
Ground state degeneracies in a many particle system may be an indication for
the presence of topological order without a local order parameter.  Well known
examples for such topological quantum liquids are the fractional Hall states
\cite{Laug83}.  Another class of systems where transitions between quantum
phases driven by topology have been proposed are two-dimensional frustrated
quantum magnets \cite{MoSo01,BaFG02,Kita06}.  Possible realizations for these
spin-liquid states are certain Iridium compounds 
\cite{magexp}.
%
As a first step towards the characterization of the different phases in a
given model
the quasi-particles of the corresponding theory need to be identified.
In a $(2+1)$ topological quantum liquid these collective excitations can be
described as defects in a planar gauge theory in a broken symmetry phase with
finite residual gauge group $H$ \cite{WiBa99}: the quasi-particles in these
systems are irreducible representations (irreps) of the Drinfeld double $D(H)$
labeled by their flux, i.e.\ an element of $h\in H$, and their topological
charge determined by the transformation properties under the residual global
symmetry commuting with the flux $h$.  States with several quasi-particles can
be manipulated by two operations: (1) fusion is determined by the
decomposition of product states into irreps of the Drinfeld double.  (2)
Interchange of two constituents in this state may reveal non-trivial anyonic
statistics corresponding to a representation of the braid group.  
Hence, representations of Drinfeld doubles allow for the description of
particles with anyonic statistics:
apart from the appearance of a phase factor, as in the case of Abelian anyons,
braiding may correspond to a unitary rotation of the wave function in a
degenerate manifold.  The fact that these non-Abelian anyons are protected by
their topological charge makes them potentially interesting as resources for
quantum computation \cite{Kita03,NSSF08}.

This picture provides a framework for the investigation of topological
phases and quantum phase transitions \cite{GTKL09}: 
%
%
Microscopic lattice models satisfying these constraints have been obtained for
the non-Abelian degrees of freedom in $su(2)_k$ Chern Simons theories,
e.g. Ising \cite{Kita06} and Fibonacci anyons 
%
\cite{LeWe05,fibo}. 
Local interactions between these objects can be varied to favour certain
fusion channels.
This allows to explore the phase diagram of these systems and to study their
critical properties near quantum phase transitions.
This approach works particularly well for quasi one-dimensional anyonic models
such as chains or ladders where powerful numerical and analytical methods are
available: at the quantum phase transition the low energy effective theory of
these systems is expected to be a conformal field theory (CFT) and the
universality class is determined by the central charge of the underlying
Virasoro algebra.  At the same time anyonic chains can be seen as realizations
for the interface between phases with different topological order
\cite{GrSc09,GATL09,FeFN09,BuSS11} and the CFT determines the properties of
gapless edge modes propagating along these interfaces.

In this letter we
study the critical behaviour of a system of interacting anyons
within an integrable quantum chain model which is constructed directly using
the algebraic structure of the gauge theory with the dihedral group of order
$6$ as its gauge group, i.e.\ the Drinfeld double $D(D_3)$. 
%
%
The general representations of the Drinfeld doubles of finite group algebras
are well known \cite{DPR90}:  
for $D(D_{3})$ one has two one-dimensional irreducible representations
$\pi_1^\pm$, four two-dimensional ones $\pi_2^{(a,b)}$, $(a,b)=(0,1)$ or
$(1,b)$ with $b=0,1,2$, and two three-dimensional ones $\pi_3^\pm$.
This leads to an increased number of fusion channels in $D(D_3)$ models as
compared to the $su(2)_k$ anyon chains, see e.g.\
Refs.~\onlinecite{Kita03,BrAC09} in the context of 2D lattice models.  On the
other hand $D(D_3)$ forms a quasi-triangular Hopf algebra which provides it
with a natural tensor product structure: the so-called quantum dimensions of
these anyons are integers.  
Therefore we can consider a chain of length $L$ with a $D(D_3)$ anyon with
three internal states represented by a copy of $\pi_{3}^{+}$ on each site.
This spin chain is a particular limit of the spin 1 Fateev-Zamolodchikov model
\cite{FaZa82} and differs from the corresponding anyonic chain in the boundary conditions.
%
For periodic boundary conditions the resulting integrable model is given by a
one parametric hamiltonian with interactions between spins on neighbouring
sites \cite{FiFL11}
\begin{equation}
\label{hamil}
  H_\theta = \sum_{i=1}^{L} \cos\theta\, h^{(1)}_{i,i+1}
                      + \sin\theta\, h^{(2)}_{i,i+1}\,. 
\end{equation}
In terms of the projection operators on irreps \cite{FDIL11} appearing in the
tensor product $\pi_3^+\otimes \pi_3^+$ the local hamiltonian can be expressed
as
\begin{equation}
\label{decomp}
  h^{(1)} = \frac{2\sqrt{3}}{3}p_1^+ - \frac{\sqrt{3}}{3}p_2^{(0,1)} -
  \frac{\sqrt{3}}{3}p_2^{(1,0)} - \frac{\sqrt{3}}{3}p_2^{(1,1)} +
  \frac{2\sqrt{3}}{3}p_2^{(1,2)}
\,.
\end{equation}
The operator $h^{(2)} = Ph^{(1)}P = \left(h^{(1)}\right)^*$ can be obtained
either by a permutation $P$ of the spins on the neighbouring sites or by
complex conjugation: with the exception of $p_2^{(1,1)} =
\left(p_2^{(1,2)}\right)^*$ the projection operators are invariant under these
operations.  Depending on the parameter $\theta$ the interactions in
(\ref{hamil}) favour different channels for the fusion of spins on
neighbouring sites, e.g.\ the vacuum channel $\pi_1^+$ for
$\pi<\theta<3\pi/2$.

By construction the local hamiltonians have the full $D(D_3)$ symmetry.  This
symmetry, however, is broken by imposing periodic boundary conditions in the
global hamiltonian (\ref{hamil}), see Ref.~\onlinecite{FiFL11}:
based on the representation theory of $D(D_3)$ the tensor product
$(\pi_3^+)^{\otimes L}$ can be decomposed into a sum of one- and
two-dimensional irreps for even $L$ (as in Eq.~(\ref{decomp}) for $L=2$).  For
the periodic chain (\ref{hamil}) only a partial decomposition of the Hilbert
space is possible.  Below we shall use this fact to assign states to (sums of)
irreps $\pi_1^+\oplus\pi_1^-$, $\pi_2^{(0,1)}$, $\pi_2^{(1,0)}$, or
$\pi_2^{(1,1)} \oplus \pi_2^{(1,2)}$.
For odd $L$ only the three-dimensional irreps $\pi_3^\pm$ appear in the tensor
product $(\pi_3^+)^{\otimes L}$.  In the spectrum of the periodic chain these
irreps cannot be separated but we find a connection between the eigenstates
and their transformation properties under the residual $D_3$ symmetry, namely
the action of the rotation $\sigma$.

For the analysis of the spectrum of the quantum chain (\ref{hamil}) we use the
property that its hamiltonian can be split into two commuting ones
with identical spectra:
%
as a consequence of the decomposition (\ref{decomp}) the local hamiltonians
$h^{(1,2)}$ commute as do the global ones (\ref{hamil}),
$[H_\theta,H_{\theta'}]=0$.  Furthermore, $H_0 \equiv \sum_ih_{i,i+1}^{(1)}$
and $H_{\pi/2} \equiv \sum_ih_{i,i+1}^{(2)}$ are related by a spatial
inversion and have the same eigenvalues albeit with opposite momentum.  These
properties can be traced back to the underlying two parameter transfer matrix
of this model and its symmetries \cite{FiFL11,Finc11}.
%
The eigenvalues of $H_0=-H_\pi$ can be parametrized by $L$ complex rapidities
$x_j$ ($\omega = \exp(2\pi i/3)$)
\begin{equation}
\label{eigen}
  E\left(X=\{x_j\}\right) = i\sum_{j=1}^{L} \frac{1}{1-i\omega\,
    \mathrm{e}^{x_{j}}} - \frac{\sqrt{3}}{3}\omega L\,
\end{equation}
solving the Bethe equations ($j=1,\ldots,L$):
\begin{equation}
\label{bae}
(-1)^{L+1} \left(
   \frac{1+(i/\omega) \mathrm{e}^{x_j}}{
         1-i\omega\, \mathrm{e}^{x_j}}\right)^{L}
  = \prod_{k=1}^{L} \frac{\mathrm{e}^{x_k} -(1/\omega)\mathrm{e}^{x_j}}{
                   \mathrm{e}^{x_k} -\omega\,\mathrm{e}^{x_j}}\,.
\end{equation}
The spectrum of the quantum chain (\ref{hamil}) is given by pairs of solutions
to these equations corresponding to energies $E_{(\alpha,\beta)}(\theta)=
\cos\theta\,E(X_\alpha)+\sin\theta\,E(X_\beta)$ and, similarly, momenta
$p=p(X_\alpha)-p(X_\beta)$, provided that the combination $(\alpha,\beta)$
satisfies the pairing rules discussed below.

In the thermodynamic limit, $L\to\infty$, all finite solutions of
Eqs.~(\ref{bae}) can be grouped into three types of so-called strings
\cite{TaSu72,FiFL11}: $\pm$-strings correspond to solutions ${Im}(x_j)=0,\pi$
and $2$-strings are complex conjugate pairs of rapidities $x_{j,\pm} =
\tilde{x}_j\pm i2\pi/3$ with real center $\tilde{x}_j$.
For finite chains this classification continues to work very well for the
ground states and low energy excitations of $\pm H_0$.  At higher energies the
$2$-strings become deformed, i.e.\ have an imaginary part different from
$\pm2\pi/3$.  We have numerically diagonalized the transfer matrix for chains
of up to $L=10$ sites and found that (taking into account this deformation)
all eigenstates can be classified this way: denoting the number of string
solutions by $N_\pm$, $N_2$ we find $N_++N_-+2N_2 = L-n_{+\infty}-n_{-\infty}$
where $n_{\pm\infty}\in\{0,1\}$ is the number of Bethe roots at $x=\pm\infty$.
%
We also find that there exist only $4\cdot 3^{L/2-1}$ ($2\cdot 3^{(L-1)/2}$)
different root configurations solving (\ref{bae}) for even (odd) length
chains.  This implies that the spectrum $H_{\theta}$ displays massive
degeneracies (exponential in $L$) arising from level crossings when $\theta$
is a multiple of $\frac{\pi}{2}$.  For generic values of $\theta$ the
degeneracies of the level $E_{(\alpha,\beta)}(\theta)$ are lifted up to a
remaining 'pairing multiplicity' of its components $\alpha$ and $\beta$.
Due to the pairing mechanism these components can be discussed separately.
For the analysis of the low energy spectrum of (\ref{hamil}) this amounts to
the identification of the ground state and low lying excitations of
$H_\pi=-H_0$ and $H_0$.
%

The ground state energy of $H_\pi$ has been computed in
Ref.~\onlinecite{FiFL11}: for even length lattices its root configuration is
given by a distribution of $N_2=L/2$ $2$-strings.
In the thermodynamic limit density functions for these strings can be
introduced allowing the energy density $\epsilon_{\infty}$ to be computed.
Here we have extended the analysis of the Bethe equations to excitations close
to this state: in the corresponding configurations one or more of the
$2$-strings are replaced by $\pm$-strings and/or Bethe roots at $\pm\infty$.
The spectrum of these excitations has a linear dispersion with Fermi velocity
$v_F$ allowing identification of the CFT for the low energy modes from the
finite size scaling behaviour of the energies for large but finite $L$
\cite{fsc}: 
%
at the critical point the ground state energy of a $1+1$ dimensional quantum
system scales as $E_0(L) - L\epsilon_\infty = -({\pi v_F}/{6L}) \,c$
where $c$ is the universal central charge of the underlying Virasoro algebra.
From the energy and momentum of low lying excitations in the finite system
\begin{equation}
\label{cft}
\begin{aligned}
  E(L) - E_0(L) = \frac{2\pi v_F}{L} \left(X+n+\bar{n}\right)\,,\quad
  p(L) - p_\infty = \frac{2\pi}{L} \left(s+n-\bar{n}\right)
\end{aligned}
\end{equation}
the scaling dimensions $X=h+\bar{h}$ and conformal spins $s = h-\bar{h}$ of
the primary fields in the theory can be determined ($n$, $\bar{n}$ are
non-negative integers).

For $H_\pi$ the product $v_{F,\pi}c$ had been determined to be $12/5$
previously \cite{FiFL11}.  Using Bethe ansatz methods and comparing the
observed structure of the low energy spectrum with (\ref{cft}) we find the
Fermi velocity for this sector to be $v_{F,\pi}= 3$.  Hence the central charge
of the effective field theory for the low energy degrees of freedom in
$H_\pi$ is $c=4/5$: this sector of the model is in the universality class of
the minimal model $\mathcal{M}_{(5,6)}$, the conformal weights $h$, $\bar{h}$
of the primary fields can take the rational values from the Kac table
%
%
$h_{pq}= ((6p-5q)^2-1)/120$, $1\le q\le p <5$.
The operator content of a given realization of the CFT is constrained further
by modular invariance of the partition function, locality of the physical
fields and boundary conditions \cite{Card86b,CaIZ87}.

In Tables \ref{tab:xpieven}, \ref{tab:xpiodd} we present numerical results for
the scaling dimensions identified in the excitation spectrum of $H_\pi$
together with the conformal predictions.
%
\begin{table}
  \caption{\label{tab:xpieven}Scaling dimensions $X_\pi$ extrapolated from the
    finite size behaviour of the ground state and low energy excitations of
    $H_\pi$ for even $L$.  $(h,\bar{h})$ are the predictions from the
    $\mathcal{M}_{(5,6)}$ minimal model.
    We have also indicated the $D(D_3)$ sector in which the state appears and
    its pairing multiplicity.  The operator content of the sector
    $\pi_2^{(1,0)}$ is obtained from that of $\pi_2^{(0,1)}$ by interchanging
    $h$ and $\bar{h}$.
  } 
\begin{ruledtabular}
 \begin{tabular}{cD{.}{.}{9}ccc}
     $D(D_3)$ & \multicolumn{1}{c}{$X_{\pi}^{\mbox{num}}$} 
              & $(h,\bar{h})$ & spin & pairing mult.
\\ \hline
     $\pi_{1}^{+} \oplus \pi_{1}^{-}$
        & 0.000000(1)&  $(0,0)$ & $0$ & 1 \\
        & 0.801(3) &  $(\frac{2}{5},\frac{2}{5})$ & $0$ & 1 \\
        & 1.80(1) & 
$(\frac{2}{5},\frac{7}{5}),(\frac{7}{5},\frac{2}{5})$ & $\pm1$ & 1 \\
\hline
     $\pi_{2}^{(0,1)}$
        & 0.4668(2) &  $(\frac{1}{15},\frac{2}{5})$ & $-\frac{1}{3}$ &
2 \\
        & 0.666666(1) &  $(\frac{2}{3},0)$ & $\frac{2}{3}$ & 2\\ \hline
     $\pi_{2}^{(1,1)} \oplus \pi_{2}^{(1,2)}$
        & 0.13334(6) &  $(\frac{1}{15},\frac{1}{15})$ & $0$ & 4 \\
        & 1.33333(3) &  $(\frac{2}{3},\frac{2}{3})$ & $0$ & 4
\end{tabular}
\end{ruledtabular}
\caption{\label{tab:xpiodd}As Table~\ref{tab:xpieven} for odd $L$.
  Symmetry is classified by the action of the $D_3$ rotation $\sigma$.}
\begin{ruledtabular}
 \begin{tabular}{cD{.}{.}{9}ccc}
     $\sigma$ & \multicolumn{1}{c}{$X_{\pi}^{\mbox{num}}$}
              & $(h,\bar{h})$ & spin & pairing mult. \\ \hline
     1
        & 0.125000(5) & $(0,\frac{1}{8})$ & $-\frac{1}{8}$ & 1 \\
        & 0.42502(2) & $(\frac{2}{5},\frac{1}{40})$ & $\frac{3}{8}$ &
1 \\
        & 0.92490(6) & $(\frac{2}{5},\frac{21}{40})$ & $-\frac{1}{8}$
& 1 \\
    & 1.625000(1) & $(0,\frac{13}{8}) $ & $-\frac{13}{8}$ & 1 \\
\hline
     $\omega$, $\omega^{-1}$
    & 0.091665(2) & $ (\frac{1}{15},\frac{1}{40})$ & $\frac{1}{24}$ & 2 \\
    & 0.59168(7) & $ (\frac{1}{15},\frac{21}{40})$ & $-\frac{11}{24}$
& 2 \\
    & 0.791667(1) & $ (\frac{2}{3},\frac{1}{8}) $ & $\frac{13}{24}$ & 2
 \end{tabular}
\end{ruledtabular}
\end{table}
For the states given we have solved the Bethe equations up to a minimum of 40
sites, although in general over 100 sites were considered when possible.  Also
listed are the residual symmetry sectors in which the levels appear.
For $L$ even the low energy spectrum of $H_\pi$ for a given symmetry coincides
with that of the 3-state Potts model subject to cyclic boundary conditions
with fixed $Z_3$ charge 
\cite{para3,Card86b} 
containing spin $\frac{1}{3}$ parafermions in the sectors $\pi_2^{(0,1)}$ and
$\pi_2^{(1,0)}$.
For $L$ odd 'twist operators' with conformal weights 
$h_{pq}$ with $q$ even appear in the (anti)holomorphic sector.  The ground
state is not invariant under a $D_3$ rotation.

Note that in spite of the appearance of fields $\Phi(h,\bar{h})$ with
conformal spin $s=h-\bar{h}\notin \mathbb{Z}/2$ physical operators in the
theory are local: they are direct products of primary fields
$\Phi(h_\alpha,\bar{h}_\alpha) \left(\Phi(h_\beta,\bar{h}_\beta)\right)^*$ in
one-to-one correspondence to the energy levels $E_{(\alpha,\beta)}$ of
(\ref{hamil}).  Momentum and spin of a physical state are given by the
difference between that of its two components.
From our numerical analysis of the spectrum we find that only operators in the
same sector with respect to the residual $D(D_3)$-symmetry (the action of the
rotation $\sigma$) pair for $L$ even (odd).  With this rule the total spin of
a physical field is either integer or half-integer.  For $-\pi<\theta<-\pi/2$
the low energy states are obtained by pairing of the states listed in Tables
\ref{tab:xpieven}, \ref{tab:xpiodd}: in the $\pi_2^{(0,1)}$ sector of the
model for even $L$ the physical fields allowed by the pairing rules carry spin
$0$ or $1$, for odd $L$ fields with spin $0,\half,1$ are possible.  Note that
the observed number of different root configurations together with the
\emph{pairing multiplicity} indicated in the tables yields the total number of
$3^L$ states of the quantum chain.


The ground state of $+H_0$ is given by a solution of the Bethe equations
(\ref{bae}) with $N_+=L/4$ $+$-strings and $N_-=3L/4$ $-$-strings
\cite{FiFL11}.  It is realized for lattices of length $L=0\pmod{4}$ and its
energy scales as $E_0(L)-L\epsilon_\infty=-3\pi/12L$ \cite{FiFL11}.
The root configurations for the lowest excitations differ from this one by the
replacement of one or two of the strings by roots at $\pm\infty$.  For
excitations at higher energies $2$-strings have to be taken into account.
Following Refs.~\cite{BoIK86,Suzu88,FrYu90,HUBBARD} the lowest finite size
energy gaps are found to be
\begin{equation}
\label{scgap2}
\begin{aligned}
  \frac{L\Delta E(\Delta N_\pm,Q_\pm) }{2\pi}
      =& \frac{1}{4}\left((\Delta N_{+})^{2} - \Delta
   N_{+} \Delta N_{-} + (\Delta 
   N_{-})^{2} \right) \\
   &+ \frac{3}{4}\left( (Q_{+})^{2} + Q_{+}  Q_{-}
   + (Q_{-})^{2} \right)\,.
\end{aligned}
\end{equation}
For the lattice model the numbers $\Delta N_\pm$ are  related to the change
in the number of $\pm$-strings as compared to the ground state, i.e.\ take
values $\mp L/4 \pmod{1}$.  They are further constrained by the fact that
the total number of Bethe roots has to be $L$.  $Q_\pm$ can take values
$Q_\pm\cong -\Delta N_\pm +\frac{1}{3}\left(n_{+\infty}-n_{-\infty}\right)
\pmod{1}$.
We can determine the Fermi velocity of low lying excitations in this sector as
before finding $v_{F,0}=3/2$.  Therefore the effective field theory for this
part of the spectrum is a CFT with central charge $c=1$.  
The field content of the theory is obtained from the finite size spectrum
(\ref{scgap2}) subject to the constraints mentioned.  It can be identified
with that of a $Z_4$ parafermionic theory \cite{ZaFa85,GeQi87}, see Tables
\ref{tab:x00mod4}, \ref{tab:x02mod4} and \ref{tab:x01mod4}.  In particular,
the finite size gap of the lowest states for $\ell=L\pmod{4}\ne0$ is
determined by an (anti-)chiral $Z_{k=4}$ spin field with conformal weight
$h_\ell= \ell(k-\ell)/(2k(k+2))$.

\begin{table}
  \caption{\label{tab:x00mod4}Scaling dimensions $X_0$ extrapolated from the
    finite size behaviour of the ground state and low energy excitations of
    $H_0$ for $L=0\pmod{4}$ (the error of the extrapolation is smaller than the
    last displayed digit).  $(h,\bar{h})$ are the predictions from the
    $Z_4$ parafermionic CFT.  For the other columns, see
    Table~\ref{tab:xpieven}. }
\begin{ruledtabular}
 \begin{tabular}{ccccc}
     $D(D_3)$ & $X_{0}^{\mbox{num}}$ & $(h,\bar{h})$ & spin & pairing mult.\\ \hline
     $\pi_{1}^{+} \oplus \pi_{1}^{-}$ 
     & $0.000000$ & $(0,0)$ & $0$ & $1$ \\ \hline 
     $\pi_{2}^{(0,1)}$
     & $0.333332$ & $(0,\frac{1}{3})$ & $-\frac{1}{3}$ & $2$ \\ \hline 
     $\pi_{2}^{(1,1)} \oplus \pi_{2}^{(1,2)}$ 
     & $0.166667$ & $(\frac{1}{12},\frac{1}{12})$ & $0$ & $4$ \\
     & $0.666667$ & $(\frac{1}{3},\frac{1}{3})$ & $0$ & $4$ \\
 \end{tabular}
\end{ruledtabular}
  \caption{\label{tab:x02mod4}As Table \ref{tab:x00mod4} but for $ L=2\pmod{4}$.}
\begin{ruledtabular}
 \begin{tabular}{ccccc}
     $D(D_3)$ & $X_{0}^{\mbox{num}}$ & $(h,\bar{h})$ & spin & pairing mult.\\ \hline
     $\pi_{1}^{+} \oplus \pi_{1}^{-}$ 
     & $0.750000$ & $(0,\frac{3}{4})\times2,(\frac{3}{4},0)\times2$ &
     $\pm\frac{3}{4}$ & $1$ \\ \hline  
     $\pi_{2}^{(0,1)}$
     & $0.083333$ & $(0,\frac{1}{12})$ & $-\frac{1}{12}$ & $2$ \\ 
     & $1.083333$ & $(\frac{3}{4},\frac{1}{3})$ & $\frac{5}{12}$ & $2$ \\ \hline 
     $\pi_{2}^{(1,1)} \oplus \pi_{2}^{(1,2)}$ 
     & $0.416667$ & $(\frac{1}{12},\frac{1}{3}),(\frac{1}{3},\frac{1}{12})$ & $\pm \frac{1}{4}$ & $4$\\
  \end{tabular}
 \end{ruledtabular}
 \caption{\label{tab:x01mod4}As Table \ref{tab:x00mod4} for $L$ odd. 
   Symmetry is classified by the action of the $D_3$ rotation $\sigma$.}
\begin{ruledtabular}
 \begin{tabular}{ccccc}
     $\sigma$ & $X_{0}^{\mbox{num}}$ & $(h,\bar{h})$ & spin & pairing mult.  \\ \hline
     $1$ 
     & $0.062500$ & $(\frac{1}{16},0)$ & $\frac{1}{16}$ & $1$ \\ 
     & $0.562500$ & $(\frac{9}{16},0)$ & $\frac{9}{16}$ & $1$ \\ 
     & $0.812500$ & $(\frac{1}{16},\frac{3}{4})$ & $-\frac{11}{16}$ & $1$ \\ \hline 
     $\omega,\omega^{-1}$
     & $0.145833$ & $(\frac{1}{16},\frac{1}{12})$ & $-\frac{1}{48}$ & $2$ \\
     & $0.395833$ & $(\frac{1}{16},\frac{1}{3})$ & $-\frac{13}{48}$ & $2$ \\
     & $0.645833$ & $(\frac{9}{16},\frac{1}{12})$ & $\frac{23}{48}$ & $2$ \\
 \end{tabular}
\end{ruledtabular}\end{table}

As in the spectrum of $H_\pi$ we find states with fractional conformal spin.
The physical fields obtained after application of the pairing rules discussed
above, however, have integer spin for $0<\theta< \pi/2$ and $L=0 \pmod{4}$.  For
lattices of length $L=2 \pmod{4}$ the spins can take integer or half-integer
values while we find $Z_4$ parafermions with quarter spin in the spectrum of
chains with $L$ odd.
For $\theta\in(\pi/2,\pi)$ or $(-\pi/2,0)$ the operators from the minimal
model $\mathcal{M}_{(5,6)}$ and the $Z_4$ CFT are paired.  In this regime
parafermionic fields with quarter spin are present for $L=2\pmod{4}$.

Both conformal field theories appearing in the low energy sector of the model
are connected with the $\widehat{sl}(2)$ affine algebra by coset
constructions.  This allows for the interpretation of the quantum chain
(\ref{hamil}) as a description for the edge of a topological fluid nucleating
within a surrounding non-Abelian liquid, similar to the interfaces between
different quantum Hall states discussed recently
\cite{GrSc09,GATL09,FeFN09,BuSS11}.
As in these models the quantum critical point is protected by topological
symmetries against local perturbations.  This becomes manifest in the small
corrections to scaling (\ref{cft}) due to deviations of the quantum chain
(\ref{hamil}) from the CFT fixed point hamiltonian: the subleading
$L$-dependence of the ground state energies indicate that these deviations are
generated by the presence of an irrelevant operator with scaling dimension
$X=3.82(3)$ in $H_\pi$ ($X=4.00(3)$ in $H_0$).

In summary, the exact solution of the model (\ref{hamil}) has allowed to
identify of the low energy effective theory for the critical phases in the
present model in terms of the direct product of two CFTs.  Thereby the model
provides examples for interfaces between different topological quantum liquids
supporting pairs of gapless modes.  The observed pairing of the CFTs indicates
the possibility of more general combinations of Virasoro characters in the low
energy spectrum than those appearing in the off-diagonal modular invariants
for the minimal model $\mathcal{M}_{(5,6)}$ or the $Z_4$ parafermionic model
alone \cite{Card86b,CaIZ87,GeQi87}.
To facilitate the explicit construction of the new invariants appearing in the
partition function of the quantum chain (\ref{hamil}) and the related $D(D_n)$
models \cite{Finc11} the spectral analysis can be extended to different
boundary conditions.  
In the context of two dimensional systems showing quantum phase transitions
between phases with different topological order, studies of these more general
models will be useful to obtain a CFT description of boundaries supporting
several edge modes and their relation to the non-Abelian bulk degrees of
freedom.

Finally, we note that for the integrable cases of braided boundaries or free
ends \cite{FiFL11} the full $D(D_n)$ symmetry is restored and the models are
equivalent to those obtained in the anyon formulation based on a fusion path
along the chain.
This equivalence can be used to embed concepts such as topological symmetry
into the framework of integrable systems allowing for a different perspective
to study possible instabilities of anyonic systems against local
perturbations.

\begin{acknowledgments}
  We thank Michael Flohr for useful discussions.  This work has been supported
  by a grant from the Deutsche Forschungsgemeinschaft.
\end{acknowledgments}
%
%
%
%

\end{document}